\documentclass[aps,twocolumn,groupedaddress,prb,showpacs]{revtex4}

\usepackage{graphicx}
\usepackage{color}

\begin{document}


\title{Kinetic theory of non-equilibrium condensation of microcavity polaritons}

\author{Davide~Sarchi}
\email[]{davide.sarchi@epfl.ch}
\affiliation{Institut de Th\'eorie des Ph\'enom\`enes Physiques, Ecole Polytechnique F\'ed\'erale de Lausanne (EPFL), CH-1015 Lausanne, Switzerland}
\author{Vincenzo~Savona}
\affiliation{Institut de Th\'eorie des Ph\'enom\`enes Physiques, Ecole Polytechnique F\'ed\'erale de Lausanne (EPFL), CH-1015 Lausanne, Switzerland}

\date{\today}

\begin{abstract}
We develop a kinetic theory of microcavity polaritons in presence
of both Coulomb and polariton-phonon interaction, obeying
particle number conservation. We study the growth of a
macroscopic population of condensed particles in the lowest
polariton state, under steady-state incoherent excitation of
higher energy states. The collective excitation spectrum,
resulting from the Coulomb Hamiltonian treated within the
Hartree-Fock-Bogolubov framework, strongly influences the
polariton condensation kinetics. In particular, for values of the
excitation intensity above the condensation threshold, scattering
from the condensate into the collective excitation modes results
in strong quantum fluctuations that deplete the condensate. A
numerical evaluation based on a few-level scheme shows that the
condensate fraction is expected to be lower than $1$ even far
above threshold. With increasing system size, the role of the
polariton quantum fluctuations becomes dominant, eventually
preventing condensation to occur for system size larger than 100
$\mu$m.
\end{abstract}
\pacs{71.36.+c,71.35.Lk,42.65.-k}

\maketitle

\section{Introduction}

Quantum fluids are the most remarkable manifestation of quantum
mechanics at the macroscopic scale. Superconductivity,
superfluidity \cite{pines66} and more recently Bose-Einstein
condensation (BEC) of diluted atoms \cite{pita03} are all
examples of a system in which many particles share the same
quantum mechanical wave function. A long sought and never
observed quantum fluid is the BEC of excitons in semiconductors.
\cite{griffin95,schmitt01} Presently, it is not well understood
why excitonic BEC eludes experimental observation. Presumably
however, three factors are believed to play against BEC. First the
structural disorder, which induces a fragmentation of the
condensate, effectively increasing the condensation critical
density as pointed out by Nozieres.~\cite{griffin95} Second, the
high rate of exciton-exciton Coulomb scattering, expected to
cause a strong dephasing of the condensate already at moderate
density,~\cite{jahnke96,ciuti00} which is not predicted by the
standard mean-field approach to
BEC.~\cite{abrik63,lozovik78,minguzzi97} Third, the strong
composite boson nature of excitons for which, contrarily to
atoms, the Mott transition density is rather close to the typical
densities for which BEC is expected to occur.~\cite{kappei05} On
the other hand, the possibility of achieving a quantum fluid in a
solid-state device, with ease of control and integration, would
open a new promising way to the implementation of quantum
information technology.~\cite{qit}

Recently, it was suggested that a quantum phase transition of
polaritons in a semiconductor microcavity under steady-state
incoherent optical pumping might occur, with formation of a
collective state of many polaritons.~
\cite{dang98,senellart99,deng02,deng03,richard05,richard05b,rubo03,laussy04,
carusotto04,carusotto05} The interest of this system resides in
the mixed nature of polaritons, which are a linear superposition
of photon and exciton states.~\cite{savona98,khitrova99} Due to
the energy-momentum dispersion of a planar microcavity, which is
parabolic around zero momentum, the resulting polariton
quasiparticle has a very light effective mass at the band bottom,
five orders of magnitude smaller than the free electron mass.
Another key feature of microcavity polaritons is the very short
radiative lifetime, spanning the range of 1 to 10 ps depending on
the quality factor of the microcavity. Because of this short
lifetime and of the much slower energy-relaxation
mechanisms,~\cite{tassone97} the polariton system under high
energy optical excitation is strongly out of thermodynamic
equilibrium. In this situation, the simple picture of an
equilibrium Bose gas is completely inadequate. Incidentally, this
also implies that the often advocated role of the light polariton
mass in determining a high BEC critical temperature, based on the
simple equilibrium expression for $T_c$, is actually irrelevant
in all experimental situations. The light mass does however play
a very important role in three other respects. First, it produces
a very long polariton coherence length which averages out the
structural disorder of the semiconductor
heterostructure,~\cite{whittaker98} thus eliminating the effect
of condensate fragmentation. Second, it reduces the final-state
phase space available for polariton-polariton scattering
processes, resulting in a strong suppression of the dephasing
compared to the exciton system,~\cite{ciuti98,savona97} which
should play in favour of BEC. Third, for an analogous reason it
is at the origin of a very slow condensation
kinetics~\cite{gardiner98} which, in the strongly nonequilibrium
condition of polaritons, could play considerably against BEC.
This complex scenario suggests that polariton BEC might be
possible provided that the density, which is required for the
condensation kinetics to be faster than the radiative lifetime,
is low enough. It also shows that a reliable theoretical model
for polariton BEC must include polariton-polariton scattering and
nonequilibrium kinetics on equal grounds.

A parallel between a microcavity polaritons collective quantum
state and conventional BEC is made hazardous by the Hohenberg-Mermin-Wagner
theorem,~\cite{hohenberg67,mermin66} stating that a phase
transition with an evident symmetry breaking is forbidden in a
2-dimensional system. For this reason, the phenomenon has been
rather interpreted as a \emph{polariton laser}
transition.~\cite{rubo03,laussy04} In more practical terms, the
BEC scenario is recovered in two dimensions if a finite system
size is considered. In this case, condensation occurs because of
the finite energy gap, due to energy quantization, between the
lowest and the first excited state of the system.~\cite{lauwers03,doan05}
This gap quenches the long-wavelength quantum fluctuations which
in the limit of infinite size destroy the condensate. For
polaritons, a finite quantization size is naturally introduced by
polariton localization over a few tens of $\mu$m resulting from
defects in the microcavity structure,~\cite{langbein02} or by the
finite size of the laser excitation spot. A few experimental
results suggest some kind of stimulated
phenomenon~\cite{dang98,senellart99,boeuf00,deng02,deng03,richard05b} or
even a phase transition with spontaneous phase
buildup,~\cite{richard05} but many observed features, among which
the unexpected observation of a thermal-type two-photon
correlation function far above threshold,~\cite{deng02} still do
not match neither the laser nor the BEC picture.~\cite{mandel95}

Laussy et al. \cite{laussy04} have pointed out that an important
role is played by the particle number conservation. Indeed, in any
symmetry breaking approach, a state with a well defined quantum
phase cannot be stationary, due to the fluctuations of the
particle number.~\cite{lewen96} Therefore, as in the theory of
BEC in diluted atoms, a number-conserving approach is needed in
order to correctly describe the quantum phase diffusion of the
condensate.~\cite{lewen96,castin98,castin01} To investigate the
appearance of condensation (either at finite temperature or in a
non-equilibrium regime), it is important to remark that in the
BEC models \cite{pines66,pita03,abrik63} both the condensed and
the non-condensed phases, having different fluctuation terms, are
considered. In these models, the interactions are the key feature
at the origin of the collective excitation spectrum, and are
responsible for the scattering kinetics that determines the ratio
between condensate and excitation
populations.~\cite{abrik63,gardiner98,leggett01} On the
theoretical side, many existing works on polaritons prefer
overlooking this aspect, pursuing a strict analogy with the laser
theory,~\cite{rubo03,laussy04} or however neglecting the role of
many-body interactions.~\cite{marchetti04}

In this work we develop a model of the polariton dynamics which
includes the polariton-polariton Coulomb interaction and the
polariton-phonon scattering on equal grounds, considering a
non-equilibrium steady-state optical pump populating the high
energy states. The model is derived within the
Hartree-Fock-Bogoliubov (HFB) approximation, as in the case of
quantum fluids at finite
temperature,~\cite{hohenberg65,griffin96} but imposing the
particle number conservation.~\cite{castin98,castin01,gardiner97}
We introduce the key assumption that the energy-relaxation
processes are much slower than the polariton-field dynamics
induced by the Coulomb interaction. This adiabatic assumption
allows to compute the relaxation kinetics onto a quasistationary
HFB spectrum. This situation, which for polaritons is justified
by the very slow relaxation kinetics within the steep region of
the lower polariton dispersion curve, is exactly opposite to the
case of BEC of diluted alkali
atoms,~\cite{barci00,gardiner98,castin97,zoller98} where the weak
mutual interactions ensure a slow field dynamics compared to the
thermalization processes. We show that the Coulomb interaction is
responsible for a depletion of the polariton condensate even far
above threshold, in favour of the excitations. This condensate
depletion is strictly related to the existence of collective
Bogolubov modes and becomes the dominant process in the limit of
large system size, effectively preventing condensation. In
Section II we derive the full theoretical formalism and discuss
its implications. Section III presents an application to a
simplified few-level model, which might describe a situation with
polariton lateral confinement and sizeable energy quantization.
Section IV is devoted to the discussion of the numerical results.
In Section V we present our conclusions and outlook.

\section{Theory}
We consider the polariton in the lower branch of the dispersion as
a quasi-particle in two dimensions, described by the Bose field $
\hat{p}_{k}$:
\begin{equation}
[\hat{p}_{k},\hat{p}^{\dagger}_{k'}]=\delta_{kk'}.
\end{equation}
The lower polariton Hamiltonian in presence of Coulomb and
polariton-phonon scattering is \cite{ciuti01,ciuti03}
\begin{eqnarray}
H&=&\sum_{k}\hbar\omega_{k}\hat{p}^{\dag}_{k}\hat{p}_{k}+\sum_{q}\hbar\omega_{q}b^{\dagger}_{q}b_{q}+H_{C}+H_{ph}\,, \label{ntotham}\\
H_{C}&=&\frac{1}{2}\sum_{kk'q}v^{(q)}_{kk'}\hat{p}^{\dag}_{k+q}\hat{p}^{\dag}_{k'-q}\hat{p}_{k'}\hat{p}_{k}\,, \label{coulint}\\
H_{ph}&=&\sum_{kk'q}g^{(q)}_{kk'}(b^{\dagger}_{q}+b_{-q})(\hat{p}^{\dag}_{k}\hat{p}_{k'}+\hat{p}^{\dag}_{k'}\hat{p}_{k}),
\label{phcoupl}
\end{eqnarray}
where the the matrix element for polariton-polariton interaction
$v^{(q)}_{kk'}$ can be derived from the Coulomb interaction
between excitons and from the oscillator strength saturation term
originating from Pauli exclusion.~\cite{ciuti01,ciuti03} The
polariton-phonon matrix element $g^{(q)}_{kk'}$ can be derived
from the deformation potential interaction with acoustic
phonons,~\cite{tassone97,doan05} which is expected to dominate at
low temperature, but could also include other electron-phonon
coupling mechanisms. {Since we aim at a kinetic description of
the polariton dynamics, we adopt the number-conserving HFB approximation. In
fact, although the total number of particles is expected to vary in presence
of a pump and of finite escape probability through the mirrors, we still
cannot lift the constraint of particle-number conservation, basically for two
reasons. First, the description of the condensate as a classical field would
result in an unphysical kinetic equation for the condensate, in which the
spontaneous in-scattering term vanishes, as shown at the end of this section.
Second, at any fixed time the number of particles is well defined in the real
system. The energy eigenvalues of the Bogolubov-like excited states depend
self-consistently on the actual number of condensed and non-condensed
particles.~\cite{gardiner97,zoller98} When including phonon relaxation, this
dependence also affects the energy relaxation rates. One way to overcome the
first of the two problems within a symmetry-breaking approach consists in
writing a separate semi-classical Boltzmann equation for the condensate
density and introducing a stimulation term scaling as the inverse of the area,
as was done by Doan {\em et al.}~\cite{doan05} Here, however, we prefer to
adopt the number-conserving formalism which directly leads to fully
self-consistent kinetic equations.}

In the
number-conserving HFB, the polariton field is expressed as
\begin{equation}
\hat{p}_k= P_{k}\hat{a}+\tilde{p}_{k},
\end{equation}
i.e. the sum of a condensate part $P_{k}\hat{a}$ and a
single-particle excitation part $\tilde{p}_{k}$~\cite{castin01}.
The condensed particle operator obeys the Bose commutation rule
$[\hat{a},\hat{a}^{\dagger}]=1$ and $N_{c}=\langle
\hat{a}^{\dagger} \hat{a} \rangle$ defines the population of
condensed particles, while $P_{k}$ represents the normalized wave
function of the condensate in momentum space. In a non-uniform
condensate~\cite{castin01}, this wave function is determined
self-consistently by imposing the relation $\langle
\hat{a}^{\dagger}\tilde{p}_{k}\rangle=0$, resulting in the
Gross-Pitaevskii equation in the thermodynamic limit. The
single-particle excitation $\tilde{p}_{k}$ is orthogonal to the
wave function of the condensate,
\begin{equation}
\sum_{k}P^{*}_{k}\tilde{p}_{k}=0,
\end{equation}
and obeys the modified Bose commutation relation
\begin{equation}
[\tilde{p}_k,\tilde{p}_{k'}^{\dagger}]=\delta_{kk'}-
P_{k}P_{k'}^{*},
\end{equation}
required to preserve the Bose commutation relation for the total
field. Using these definitions, the total population of particles
with momentum $k$ is
\begin{equation}
N_{k}=\langle \hat{p}^{\dagger}_{k}\hat{p}_{k}\rangle =|P
_{k}|^{2} N_{c} + \tilde{N}_{k},
\end{equation}
where $\tilde{N}_{k}=\langle
\tilde{p}^{\dagger}_{k}\tilde{p}_{k}\rangle$ is the non-condensed
population.

The time evolution of the populations can be evaluated by means of
the Heisenberg equations of motion. As a first step, we consider
only the Coulomb interaction Hamiltonian, neglecting the
polariton-phonon scattering. We then obtain the following
equations for the dynamics of the field operators:
\begin{equation}
i\dot{\hat{a}}=\sum_k i \dot{P}^*_k \hat{p}_k+\sum_k P^*_k
(\omega_k \hat{p}_k + v \sum_{k',q} \hat{p}^{\dagger}_{k'-q}
\hat{p}_{k'} \hat{p}_{k-q} ) \label{eq:aoper}
\end{equation}
and
\begin{equation}
i\dot{\tilde{p}}_k= \omega_k \hat{p}_k + v \sum_{k',q}
\hat{p}^{\dagger}_{k'-q} \hat{p}_{k'} \hat{p}_{k-q} - i \dot{P}_k
\hat{a} - i P_k \dot{\hat{a}}. \label{eq:ptoper}
\end{equation}
Notice that here and in the following we assume a contact
polariton-polariton interaction, i.e. $v^{q}_{kk'}\equiv \hbar v$,
and we adopt mean-field factorizations, as detailed in the
Appendix. Let us introduce the amplitude of the scattering
process bringing two particles from the non-condensate to the
condensate
\begin{equation}
\tilde{m}_{k,k'}\equiv \langle
\hat{a}^{\dagger}\hat{a}^{\dagger}\tilde{p}_k\tilde{p}_{k'}\rangle.
\label{def:m}
\end{equation}
Then the Heisenberg equations result in the following time
evolution of the condensate population
\begin{eqnarray}
\dot{N}_c=-\sum_{k}\dot{\tilde{N}}_k&=& 2 v {\mathop{\rm
Im}\nolimits} \left\{ \sum_{k,k',q} P^*_k \langle
\hat{a}^{\dagger}\hat{p}^{\dagger}_{k'-q} \hat{p}_{k'}
\hat{p}_{k-q} \rangle \right\} \nonumber \\
&=& 2 v {\mathop{\rm Im}\nolimits} \left\{ \sum_{k,k',q} P^*_k
P^*_{k'-q} \tilde{m}_{k',k-q} \right\}. \label{eq:nc}
\end{eqnarray}

In order to simplify the present analysis, we specialize the
model to a spatially homogeneous system. In such a limit, the
condensate wave function can be safely assumed as a homogeneous
function in the spatial domain. We expect the total deviation
from this approximation to be small, as the actual wave-function
will differ only at the system boundaries. In momentum space,
this assumption implies $P_{k}=e^{i\phi} \delta_{0,k}$, where
$\phi$ is the condensate macroscopic phase. The assumption
therefore implies that the condensate state is always
characterized by $k=0$. In the following the macroscopic phase
factor $e^{i\phi}$, together with its time dependence, will be
included in the definition of the operator $\hat{a}$. We point
out that the assumption of a homogeneous system is not in
contrast with that of finite size, provided the system size is
large enough to neglect boundary effects. The state orthogonality
in this case implies $\hat{p}_k=\tilde{p}_k$ for $k \ne 0$. As a
result, only the diagonal scattering amplitudes
$\tilde{m}_k=\langle
\hat{a}^{\dagger}\hat{a}^{\dagger}\tilde{p}_k\tilde{p}_{-k}\rangle$
appear in the equations, in analogy with the anomalous
correlations entering the standard HFB approach.~\cite{griffin96}
In this limit, Eq. (\ref{eq:nc}) becomes
\begin{equation}
\dot{N}_c = 2 v {\mathop{\rm Im}\nolimits} \left\{
\sum_k\tilde{m}_{k}\right\}\equiv 2 v {\mathop{\rm Im}\nolimits}
\left\{\tilde{m}\right\}, \label{eq:nc2}
\end{equation}
where we have defined the total scattering amplitude $\tilde{m}$
as a sum over all possible final states of the Coulomb scattering
process.

Turning to the kinetics of the non-condensate degrees of freedom,
we define the destruction operator of a condensate
excitation~\cite{castin98}
\begin{equation}
\hat{\Lambda}_k\equiv\frac{1}{\sqrt{N}}\hat{a}^{\dagger}\tilde{p}_k\,,
\end{equation}
that creates a condensate particle by destroying a non-condensate
one. This operator obeys quasi-Bose commutation rules
\begin{equation}
\left[\hat{\Lambda}_k,\hat{\Lambda}_q\right]=0 \label{eq:comlamb1}
\end{equation}
and
\begin{equation}
\left[\hat{\Lambda}_k,\hat{\Lambda}^{\dagger}_q\right]=\delta_{k,q}(N_c-\tilde{N}_k-1)/N.
\label{eq:comlamb2}
\end{equation}
Introducing the standard Bogolubov transformation, the
single-particle excitations can be expressed as
\begin{equation} \hat{\Lambda}_k=U_k\hat{\alpha}_k+V^*_{-k}\hat{\alpha}^{\dagger}_{-k},
\label{eq:bogtr}
\end{equation}
where $U_k$ and $V^*_{-k}$ are modal functions, and
$\hat{\alpha}_k$ are the operators for Bose normal modes
corresponding to the collective excitations of the
system.~\cite{pita03,castin98} In particular, the commutation
rule (\ref{eq:comlamb1}) impose the condition $U_k V^*_k=U_{-k}
V^{*}_{-k}$.~\cite{castin98,note1} Using Eq. (\ref{eq:bogtr}), we
obtain a direct relation between the one-particle density matrix
$\langle \tilde{p}^{\dagger}_k \tilde{p}_{k'} \rangle$ and the
populations of Bogolubov modes $\bar{N}_k=\langle
\hat{\alpha}^{\dagger}_k \hat{\alpha}_k \rangle$:
\begin{equation}
\langle \tilde{p}^{\dagger}_k \tilde{p}_{k'} \rangle \sim
\delta_{kk'} [(|U_k|^2 + |V_k|^2)\bar{N}_k + |V_k|^2].
\end{equation}
This brings to the result, expected by symmetry arguments, that
for a spatially homogeneous system the off-diagonal density
matrix terms of the non-condensed states are vanishing within the
mean-field approach, i.e.
\begin{equation}
\tilde{N}_{k,k'}\equiv \langle
\tilde{p}^{\dagger}_k\tilde{p}_{k'}\rangle=\tilde{N}_{k}\delta_{k,k'}.
\label{eq:offdiag}
\end{equation}
Hence, we obtain the following equation for the population of the
excited states
\begin{equation}
\dot{\tilde N}_{k} = - 2v {\mathop{\rm Im}\nolimits}\{\tilde{m}_k
- \sum_{q} \langle
\tilde{p}^{\dagger}_{q}\tilde{p}^{\dagger}_{-q}\tilde{p}_{k}\tilde{p}_{-k}\rangle\}.
\label{eq:Ntk}
\end{equation}
Finally the scattering amplitude $\tilde{m}_k$ obeys the equation
(see the Appendix for the derivation)
\begin{eqnarray}
\dot{\tilde m}_k &=& i \Omega_k \tilde m_{k}-2 i v \tilde{N}_k \tilde{m}- i v (1 + 2 \tilde{N}_{k}) N_{c} (N_{c}-1)\nonumber \\
&+& i v (1 + 2 N_{c})\sum_{q}\langle \tilde{p}^{\dagger}_{q}
\tilde{p}^{\dagger}_{-q} \tilde{p}_{k} \tilde{p}_{-k} \rangle\,.
\label{eq:coulomb}
\end{eqnarray}
where $\Omega_k=-2[\omega_k+v(N_c-\tilde{N}_k-5/2)]$.

The residual two-particle correlations for excited particles,
appearing in Eqs. (\ref{eq:Ntk}) and (\ref{eq:coulomb}), require
special care as they are the source of the off-diagonal long-range
correlations characterizing a Bose-Einstein condensate in real
space. Once again, these terms in the number-conserving approach
are analogous to the anomalous correlations appearing in a
standard HFB formalism.~\cite{griffin96} In particular, they
cannot be factored in the single-particle basis without affecting
the spatial correlation properties of the condensate. We will see
in the next section how these terms can be treated within a
simplified few-level model. Notice that the quantity $\tilde{m}$
denotes a scattering process which would not conserve energy in a
single-particle picture, as it describes the scattering of two
particles from one state to another one at larger energy. This
process is actually present in our formalism because the spectrum
of the system is modified by the interactions and the new
eigenstates are the collective Bogolubov excitations, describing
condensate fluctuations with large wavelength, rather than the
single-particle states.~\cite{pita03}

We now introduce the contribution to the population kinetics due to polariton-phonon scattering.
In the limit of a spatially
homogeneous system, we can write the phonon contribution to
the population kinetics of the condensate as
\begin{equation}
\left.\dot{N}_c\right|_{ph} = -\frac{i}{\hbar} \sum_{q,k}
g^{(q)}_{k0}\langle
(b^{\dagger}_{q}+b_{-q})(\hat{a}^{\dag}\tilde{p}_{k}-\tilde{p}^{\dag}_{k}\hat{a})
\rangle.
\end{equation}
The Heisenberg equations for the operators produce again a
hierarchy of equations for phonon-assisted correlations of all
orders, which are coupled to the HFB variables. The equations for
the first order phonon-assisted correlations, like e.g. $\langle
b^{\dagger}_q \hat{a}^{\dagger} \tilde{p}_{k}\rangle$, are
formally solved within the self-consistent Markov approximation.
In particular, the higher order phonon assisted correlations
entering the kinetic equations are factored according to the mean
field approximation. For example
\begin{equation}
\langle\hat{a}^{\dagger} \hat{a}^{\dagger} \hat{a}
b_{q}^{\dagger}\tilde{p}_{k} \rangle \simeq (N_c-1) \langle
b^{\dagger}_q \hat{a}^{\dagger} \tilde{p}_{k}\rangle
\end{equation}
or
\begin{equation}
\langle b_{-q'} b_{q}^{\dagger} \tilde{p}_{k'}^{\dagger}
\tilde{p}_{k}
 \rangle \simeq \delta_{q,q'}\delta_{k,k'}\tilde{N}_k (1+n_q),
\end{equation}
where $n_q$ is the phonon distribution at wave vector $q$. In
this way, we obtain equations of the following kind
\begin{eqnarray}
i\partial_t \langle b^{\dagger}_q \hat{a}^{\dagger}
\tilde{p}_{k}\rangle &=& \left[ \omega_k + v(N_c+\tilde{N}_k -1)
-w_q \right]\langle b^{\dagger}_q \hat{a}^{\dagger}
\tilde{p}_{k}\rangle \nonumber \\
&+& \frac{g^{(q)}_{k,0}}{\hbar}\left[(n_q-\tilde{N}_k)N_c
-(1+n_q)\tilde{N}_k \right], \label{eq:1orph}
\end{eqnarray}
whose formal solution can be plugged into the HFB equations. The
resulting contribution to the dynamics of the condensate
population is:
\begin{eqnarray}
\left.\dot{N}_c\right|_{ph} &=& 2 \pi \sum_{q,k} \delta
\left(\omega_k +
v(N_c+\tilde{N}_k -1) -w_q \right) \label{eq:markov}\\
& \times & \left|\frac{g^{(q)}_{k0}}{\hbar}\right|^2
\left[(\tilde{N}_k - n_q) N_c +
(1+n_q)\tilde{N}_k \right] \nonumber \\
& \equiv & 2 \sum_k g_c^{(k)} \left[(\tilde{N}_k - n_c^{(k)}) N_c
+ (1+n_c^{(k)})\tilde{N}_k \right], \nonumber
\end{eqnarray}
where $g_c^{(k)}$ and $n_c^{(k)}$ are respectively the effective
phonon scattering matrix element and the phonon population at the
wave vector defined by momentum and energy conservation. In
particular, the in-plane wave vector component $q_{\parallel}$ is
fixed by momentum conservation, while the z-component $q_z$ is
selected by energy conservation. In this way, the
polariton-phonon coupling introduces effective phonon-mediated
polariton-polariton interaction terms,~\cite{zimmermann03} the
lowest-order ones being proportional to $|g^{(q)}_{kk'}|^2$. The
phonon populations $n_q$ are assumed to be thermally distributed
at the lattice temperature. Eq. (\ref{eq:markov}) is the standard
Boltzmann equation expected for the energy-relaxation kinetics. A
similar Boltzmann equation holds for the populations
$\tilde{N}_k$ of the excited single-particle states. The
phonon-assisted correlations entering the equations for the
scattering amplitudes $\tilde{m}_k$, on the other hand, cannot be
solved analytically as was done for Eq. (\ref{eq:1orph}), because
Coulomb interaction couples different values of the momentum.

{We conclude this section by giving the explicit
expression of the phonon coupling term, as it would appear in the
equation for the condensate if a symmetry-breaking approach was adopted.
Expressing the polariton field as
\begin{equation}
\hat{p}_k=P_k+\tilde{p}_k,
 \label{eq:standardhfb}
\end{equation}
where $P_k=\langle\hat{p}_k\rangle$ is a classical field, while
$\tilde{p}_k$ describes fluctuations, and considering again the
uniform limit $P_k=\delta_{k,0}P$, the phonon contribution to the condensate field equation reads
\begin{equation}
\left.\dot{P}\right|_{ph} = -\frac{i}{\hbar} \sum_{q,k}
g^{(q)}_{k0}\langle (b^{\dagger}_{q}+b_{-q})\tilde{p}_{k} \rangle.
\end{equation}
Following the same
procedure adopted above in order to evaluate the phonon assisted
terms within the Markov approximation, we obtain for the
condensate population $N_c \equiv |P|^{2}$ the following
expression
\begin{equation}
\left.\dot{N}_c\right|_{ph} = 2\pi \sum_{q,k} \delta
\left(\omega_k - w_q \right)
\left|\frac{g^{(q)}_{k0}}{\hbar}\right|^2 (\tilde{N}_k - n_q) N_c,
\end{equation}
Where the in-scattering term present in Eq.
(\ref{eq:markov}) is here instead missing. This term, responsible for the
spontaneous scattering into the condensate, is of course absent
within a description of the condensate as a classical field.
As suggested at the beginning of this section, the adoption of
the number-conserving formalism prevents the occurrence of this unphysical behavior.}

\section{Few-level model}
\begin{figure}[h!]
\includegraphics[width=.47 \textwidth]{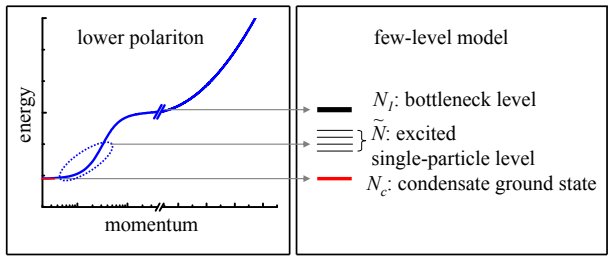}
\caption{\label{fig1}Schematic lower polariton dispersion. The
relaxation from the bottleneck region to the ground state is
described by a few-level model in which a bottleneck level, a
single-particle excited level and the ground level are
introduced.}
\end{figure}

As seen in the previous Section, the
solution of the whole set of equations for the populations and
correlations is a challenging task, basically because of the
off-diagonal coupling in the phonon assisted correlations. In
addition, the two-point correlations between single-particle
excitations appearing in Eqs. (\ref{eq:Ntk}) and
(\ref{eq:coulomb}) still need to be addressed in a consistent
way. In a typical photoluminescence experiment under nonresonant
excitation, the steep polariton dispersion results in a
relaxation bottleneck,~\cite{doan02,tassone97} with polariton
population piling up at the boundary of the flat exciton-like
region of the polariton dispersion. From there, polaritons relax
to the band-bottom, where the actual phase transition can take
place before they recombine emitting a photon. Given the very
slow relaxation rates~\cite{tassone97} and the very fast
radiative recombination rates, the relaxation from the bottleneck
region to the polariton band bottom is very likely to take place
in one, or at most a few relaxation steps. Multiple relaxation
steps within the steep region of the polariton dispersion are
however very unlikely. Their contribution might quantitatively
affect the total out-scattering rate from the bottleneck region,
but it will not affect in a sizeable way the in-scattering rate
in the ground level, which is the relevant process for
final-state stimulation. This allows the introduction of an
effective-level scheme, which should describe the relevant
dynamics of the problem. In this scheme, the flat bottleneck
region of the dispersion, the excited states within the steep
region and the ground state are described as three effective
levels, accounting for the respective density of states in the
real system. We will refer to these levels as {\em bottleneck},
{\em single-particle excitations} and {\em condensate}
respectively. For the reasons illustrated above, we can assume
that the phonon mediated relaxation occurs only between the
bottleneck level and the two other levels, while the Coulomb
amplitudes $\tilde{m}_k$ are non-vanishing only for states
energetically close to the condensate, i.e. in the band bottom
region. The first assumption, as already stated, stems from the
fast radiative rate of the polariton levels in the strong
coupling region of the dispersion, ensuring a negligible
contribution to the in-scattering rate in the ground
level.~\cite{tassone97} The second assumption is justified by the
fact that the Coulomb scattering amplitudes $\tilde{m}_k$ can be
important only for small wave vector $k$, because they are linked
to condensate fluctuation of large wavelength, as discussed
previously. This scheme is analogous to the one commonly adopted
for the description of BEC kinetics of a diluted alkali
gas.~\cite{zoller98} The idea behind it is that in the lower
energy region of the spectrum, the Bogolubov field dynamics is
the dominant process, while the low density of states makes the
relaxation kinetics negligible. The opposite occurs in the
higher-energy region, where most of the relaxation kinetics takes
place but the collective Bogolubov excitations coincide with the
single-particle states.~\cite{dalfovo99} Our simplified few-level
model is sketched in Fig. \ref{fig1}. The creation operator for
the bottleneck state is defined by $\hat{p}^{\dagger}_{1}$ (and
$N_1$ is the population per mode at bottleneck), while
$\tilde{p}^{\dagger}$ now indicates the creator for a
single-particle excitation. The corresponding total population of
single-particle excitations is given by
$\tilde{N}=\sum_{k}\tilde{N}_k$. Our purpose is to write a set of
equations giving the time evolution of the condensate population
$N_c$, of the single-particle excitations population $\tilde{N}$
and of the bottleneck population $N_1$.

As in previous treatments,~\cite{porras02} the introduction of
effective levels requires the introduction of renormalized
coupling constants. In our case, the effective phonon coupling
rates involving the bottleneck level are proportional to the
total number of states $\rho_x$ in this region of the dispersion.
This latter is estimated from the assumption of a thermalized
polariton distribution at the bottleneck,~\cite{porras02} and is
related to the exciton mass $M_{exc}$ and the exciton energy
thermal broadening $E\approx k_{B}T$ by
\begin{equation}
\rho _{x}  =(A/2\pi)(M_{exc}E/\hbar^{2}), \label{eq:rhox}
\end{equation}
resulting in an effective phonon coupling rate $g_k=\rho_x
g^{(1)}_k$, where $g^{(1)}_k$ is the phonon coupling rate for a
scattering process between a state at the bottleneck and a
polariton state with momentum $k$, defined as in equation
(\ref{eq:markov}).~\cite{tassone97}
We now address the problem of the two-particle correlation terms
entering equations (\ref{eq:Ntk}) and (\ref{eq:coulomb}). Using
again Bogolubov transformations (\ref{eq:bogtr}), linking the
single-particle field to collective excitations, we can write
{
\begin{eqnarray}
& & \sum_{q,k}\langle
\tilde{p}^{\dagger}_{q}\tilde{p}^{\dagger}_{-q}\tilde{p}_{k}\tilde{p}_{-k}\rangle
=  \Upsilon(N) \left[\left|\sum_k U_k V^*_k(1+2\bar{N}_k)\right|^2\right. \nonumber \\
& & + \left. \sum_k 2\chi_k \bar{N}_k \left(\chi_k
\bar{N}_k+2\left| V_k \right|^2\right)+2\left| V_k \right|^4
\right], \label{eq:2corr_bog}
\end{eqnarray}
where $\Upsilon(N)=N^2[(N_c+1)(N_c+2)]^{-1}$ and
$\chi_k=\xi_k+2\left| V_k \right|^2$. The quantity
$\xi_k=(N_c-\tilde{N}_k-1)/N$ comes from the normalization of the
Bogolubov factors $|U_k|^2-|V_k|^2=\xi_k$.~\cite{note1} One of
the key points of the present approach consists in evaluating the
Bogolubov coefficients self-consistently as a function of the
instantaneous populations at each time in the kinetic. This
amounts to replacing in Eq. (\ref{eq:2corr_bog}), the expression
\begin{equation}
\left|V_k\right|^2=\xi_k\frac{\left[E_k-(\omega_k+\sigma_k)\right]^2}{\sigma^2_k-\left[E_k-(\omega_k+\sigma_k)\right]^2},
\label{eq:v_bog}
\end{equation}
where
\begin{equation}
E_k=\left[(\omega_k+\sigma_k)^2-\sigma_k^2\right]^{1/2}
\label{eq:E_bog}
\end{equation}
is the Bogolubov quasi-particle energy and $\sigma_k=v N \xi_k$.
The physical interpretation of this approach is that the
relaxation kinetics is much slower than the Bogolubov field
dynamics. We are therefore assuming a quasi-stationary situation
where, at each time during the evolution of the system, a
Bogolubov spectrum can be evaluated from the instantaneous
distribution of the quasiparticle populations. This is of course
only allowed in the limit of validity of the number-conserving
Bogolubov approach, namely for $N_c >
\tilde{N}_k$.\cite{castin98}}

{The terms in Eq.(\ref{eq:2corr_bog}) still contain
k-dependent factors but we can see that, for $\tilde{N}>1$, the
dominant contribution is proportional to $\tilde{N}^2$, via the
relation
\begin{equation}
\bar{N}_k=\left[(1+N_c)\tilde{N}_k-\left|V_k\right|^2\right]\chi_k^{-1}.
\label{eq:Nbog_nsp}
\end{equation}
The amplitude of this term depends on the anomalous coefficients
$V_k$, which tend to zero for vanishing condensate density.
Therefore, in the thermodynamic limit, since the quantities $N_c$
and $V_k$ vanishes as expected for a Bose system in two
dimensions, the quantity in Eq.(\ref{eq:2corr_bog}) reduces to a
term proportional to $\tilde{n} \tilde{N}$, where $\tilde{n}$ is
the particle density.} On the other hand, in order to solve the
effective-level equations presented below, we want rewrite this
contribution in a more useful k-independent form. To this
purpose, we propose the following argument based on the two-point
spatial correlation function. In real space, we can write
\begin{equation}
\sum_{q,k}\langle
\tilde{p}^{\dagger}_{q}\tilde{p}^{\dagger}_{-q}\tilde{p}_{k}\tilde{p}_{-k}\rangle
= \int \,d{\bf r}d{\bf s}\langle\tilde{p}^{\dagger}({\bf
r})\tilde{p}^{\dagger}({\bf r})\tilde{p}({\bf s})\tilde{p}({\bf
s})\rangle.
\label{eq:realsp}
\end{equation}
Now, {for a non-condensed system,} at the lowest order in the total density, the four-point
spatial correlation can be {safely} factored in terms of the two-point
correlation $\langle \tilde{p}^{\dagger}({\bf r})\tilde{p}({\bf
s})\rangle$ as
\begin{equation}
\langle\tilde{p}^{\dagger}({\bf r})\tilde{p}^{\dagger}({\bf
r})\tilde{p}({\bf s})\tilde{p}({\bf s})\rangle \simeq
2\langle\tilde{p}^{\dagger}({\bf r})\tilde{p}({\bf
s})\rangle^2-\delta({\bf r}-{\bf s})\tilde{N}({\bf r}).
\label{eq:fac2pt}
\end{equation}
{Therefore,} defining the parameter
\begin{equation}
\alpha\equiv \tilde{N}^{-2}\int d{\bf r}d{\bf
s}\langle\tilde{p}^{\dagger}({\bf r})\tilde{p}({\bf s})\rangle^2,
\label{eq:def_a}
\end{equation}
we obtain
\begin{equation}
\sum_{q,k}\langle
\tilde{p}^{\dagger}_{q}\tilde{p}^{\dagger}_{-q}\tilde{p}_{k}\tilde{p}_{-k}\rangle
\simeq (2\alpha\tilde{N}-1)\tilde{N}, \label{eq:2Bcor}
\end{equation}
{In an equilibrium situation, for a non-condensed
phase, the two-point spatial correlation vanishes at distances
larger than the thermal length
$\lambda_T=\sqrt{2\pi\hbar^2/(mk_BT)}$, which derives directly
from the equilibrium Bose-Einstein distribution of excitations.
In a non-equilibrium situation like the present one, the
distribution of excitations generally differs from the
equilibrium one, but again it corresponds to a characteristic
length. We however expect the correlation length $\kappa$ of the
one-body density function, defining the spatial correlation
length for a non-condensed system, to be in the same range as
$\lambda_T$. Therefore, from Eq. (\ref{eq:def_a}) we see that
\begin{equation}
\alpha \sim \frac{\kappa^2}{A}.
\label{eq:alfa_unp}
\end{equation}
Assuming $\kappa$ independent of the system size, $\alpha$ then
scales as $A^{-1}$.~\cite{pita03} For a system that has undergone
condensation, on the other hand, the simple factorization used in
Eq.(\ref{eq:fac2pt}) is not valid, as it totally neglects the
anomalous correlations. These contributions are the dominant term
in Eq.(\ref{eq:2corr_bog}), proportional to the products $U_k
V^*_k$. Inspection of Eq. (\ref{eq:2corr_bog}), shows that its
dominant term still depends on the squared population of
excitations $\tilde{N}^2$, as previously discussed. We therefore
propose to use the expression (\ref{eq:2Bcor}) also in the
condensate regime, by making a different assumption on the
parameter $\alpha$. In particular, for a condensed system, the
parameter $\alpha$ should no longer scale as in Eq.
(\ref{eq:alfa_unp}), because of the presence of long-range
correlations. At the same time, comparing equations
(\ref{eq:realsp}) and (\ref{eq:2Bcor}), we see that only values
$\alpha<1$ are admitted, the limiting value $\alpha=1$
corresponding to a four-point spatial correlation function
extending over the whole system size. We still expect, however, a
residual dependence of $\alpha$ on the system area $A$ even in
the condensed regime, because by increasing the system size we
suppress the relaxation mechanism in favour of the condensate
depletion. Therefore, a self-consistent evaluation of the
parameter $\alpha$ is needed for the present kinetic model to
have the correct thermodynamic limit. We discuss this approach
below, after having introduced the effective-level description of
the system.} Concluding this analysis, we remark that the quantity
(\ref{eq:2Bcor}) is real, so the last term in equation
(\ref{eq:Ntk}) gives a vanishing contribution when summed over
$k$.

In order to obtain a closed set of equations for $N_{c}$,
$\tilde{N}$, $N_{1}$ and $\tilde{m}$, the energies of the
single-particle excitations are replaced by an effective value,
$\hbar\bar{\omega}\simeq\hbar\omega_k$ representing a typical
energy of the non-condensed levels relative to the ground state. This quantity plays a crucial role in this effective-level scheme, as it determines the finite energy gap that makes condensation possible in a two dimensional system. We discuss below how the numerical value of this parameter is chosen.
Similarly, we introduce the energy $\hbar\omega_1$ for the
bottleneck states. We assume intrinsic linewidths
$\hbar\gamma_{1}$ for the bottleneck level and $\hbar\gamma_{0}$
for the other levels, accounting for radiative recombination as
well as nonradiative homogeneous energy broadening. In this way,
using the relation (\ref{eq:2Bcor}), the resulting equations are
\begin{widetext}
\begin{eqnarray}
\dot{N}_c  &=&  - 2 \gamma_0 N_{c} + 2 g_c \Gamma
(1+n_c)N_1(1+N_c)- 2 g_c \Gamma n_c  (1+N_{1})N_c + 2 v{\mathop{\rm Im}\nolimits} \{\tilde{m}\}\,, \label{eqall}\\
\dot{\tilde N}  &=&  - 2 \gamma_0 \tilde N + 2 \tilde g \Gamma (1+\tilde n)N_1(\eta+\tilde N) - 2 \tilde g \Gamma \tilde n (1+N_{1})\tilde N - 2 v  {\mathop{\rm Im}\nolimits}\{\tilde{m}\}\,, \nonumber\\
\dot{N}_{1}  &=&  - 2 \gamma_1 N_{1} + 2 \Gamma (g_c n_c N_c +
\tilde g \tilde n \tilde N )(1+N_{1})\nonumber \\
&& - 2 \Gamma  \left[ g_c(1+ n_c)(1+N_c)+\tilde g (1+\tilde n)(\eta+\tilde N) \right] N_1 + F\,, \nonumber\\
\dot{\tilde m} &=&  - 2 \left\{2\gamma_0+
\Gamma_{m}\left[g^{\prime} (n^{\prime} - N_1)+g^{\prime \prime}
(n^{\prime \prime} - N_1)\right] + i \Omega' \right\} \tilde m
\nonumber \\
&& + i v \left[(1+2N_{c})\tilde{N}(2\alpha \tilde{N}-1) -
(\eta+2\tilde{N})N_{c}(N_{c}-1)\right]\,, \nonumber
\end{eqnarray}
\end{widetext}
where $\Gamma=\gamma_{1}/(\gamma_{1}+\gamma_{0})$,
$\Gamma_{m}=\gamma_{1}/(\gamma_{1}+3\gamma_{0})$,
$\Omega'=[\bar{\omega}+v(N_c+(1-1/\eta)\tilde{N}-5/2)]$ and
$\eta$ is the number of single-particle excitation states,
depending on the quantization area via the relation $k=(n_x{\bf
x} + n_y {\bf y})\pi/\sqrt{A}$.~\cite{porras02,tassone00} We have
denoted by $n_c$, $\tilde n$, $n^{\prime}$ and $n^{\prime
\prime}$ the phonon populations defined by energy conservation
within the Markov assumption, as shown in equation
(\ref{eq:markov}). Correspondingly, $g_c$, $\tilde g$,
$g^{\prime}$ and $g^{\prime \prime}$ denote the effective phonon
coupling strength at the same energies, renormalized by the
number of bottleneck levels $\rho_x$. The equation for the
scattering amplitude contains an oscillating term, whose
frequency depends on the actual energy needed to create
condensate particle. Notice that all this quantities depend on
the actual condensate and non-condensate densities and vary
self-consistently during time evolution.
{In particular, the parameter $\alpha$ can be obtained self-consistently at each step of the kinetics, by equating expressions (\ref{eq:2corr_bog}) and (\ref{eq:2Bcor}).
The effective-level representation of the condensate excitation naturally implies an effective Bogolubov factor $V$, via Eq. (\ref{eq:v_bog}), and corresponding values for
the other related quantities. The expression for $\alpha$ takes then the compact
form
\begin{equation}
\alpha(N)=\frac{1+N_c}{2+N_c}\left(\frac{1}{\eta}+2\frac{\left|U^*V\right|^2}{\chi^2}\right)\,.
\label{eq:alfa_sc}
\end{equation}
In particular,
we find again that the first term in parenthesis scales as
$A^{-1}$, consistently with the previous discussion. Actually, it turns out that the precise value of
$\alpha$ is not critical in determining the condensation
dynamics, provided that the two limits $\alpha=0$ and $\alpha=1$
are not reached. In particular, in the next section, we will show
how the self-consistent result does not differ much by one one
obtained using a constant value for $\alpha$.
We have introduced a steady-state pump rate $F$ in the equation for the
effective bottleneck level. This quantity represents the number of particles
per unit time and per state, which enters the system following the relaxation
from higher-energy states, as in a typical experiment with non-resonant
continuous-wave excitation.~\cite{deng02,deng03,dang98,senellart99} For a
given total pump flux $f$ (total rate of polaritons per unit area), the
quantity $F$ can be rewritten as $F=f A \rho_{x}^{-1}$. This shows that for a
given pump flux the quantity $F$ does not depend on the area $A$ of the
system, as can be argued from Eq. (\ref{eq:rhox}).}

The phonon-mediated interaction results in a Boltzmann
dynamics,~\cite{tassone97} with scattering coefficients
$g_c\Gamma$, $\tilde{g}\Gamma$, $g^{\prime}\Gamma_m$ and
$g^{\prime \prime}\Gamma_m$. In the kinetics described by our
model, the initial growth of the condensate is triggered by
final-state stimulation within the relaxation process. It takes
place in the ground state which is separated by an energy gap
$\bar{\omega}$ from the excited level. Above the stimulation
threshold, a crucial role is played by the Coulomb scattering
amplitude $\tilde{m}$, which enters the first two equations in
(\ref{eqall}) with equal factors up to a sign, and determines the
actual condensate growth or depletion. As we will see, this term
determines the macroscopic condensate fraction reached by the
system. Indeed, we can compute the approximate analytical steady
state fraction, neglecting the oscillating term in the equation
for $\tilde{m}$ and making the high density limit $\tilde{N}\gg
1$,$N_c\gg 1$ with $N_1/N_c\rightarrow 0$. This is justified by
the fact that, above threshold, $N_1$ is expected to reach a fixed
value $N_1^{(thr)}\sim F_{th}\gamma_1^{-1}$. Neglecting the phonon population and taking
$\Gamma_m g^{\prime} \simeq \Gamma_m g^{\prime \prime}\equiv g_m$
and $\Gamma g_c\equiv \bar{g}_c$, we obtain for $\tilde{m}$ the
steady state solution
\begin{equation}
\tilde{m}=i \frac{v}{2}
\frac{(2\alpha\tilde{N}-N_c)N_c\tilde{N}}{\gamma_0-g_m
F_{th}\gamma_1^{-1}}.
\end{equation}
This result, plugged into equation for $N_c$, gives
\begin{equation}
N_c \simeq 2\alpha\tilde{N}-2\frac{(\gamma_0-g_c
F_{th}\gamma_1^{-1})(\gamma_0-g_m F_{th}\gamma_1^{-1})}{v^2\tilde{N}^2},
\end{equation}
predicting the coexistence of condensate and non-condensate even
far above threshold. This behaviour, as will be also seen in the numerical solution,
originates exclusively from the Coulomb scattering, whereas a pure Boltzmann dynamics
would always predict a condensate fraction approaching 1 above threshold.

{The adoption of this simplified model calls for some
additional remarks, in order to understand its limitations and the relevance of the
parameters. As already argued, we can expect that
the relaxation from the bottleneck region into the
low-energy states is qualitative well described within a few-level approximation, as the inclusion of other
intermediate states would only
result in a finite increase of the pump threshold.
This in turn would not introduce significant changes to the
exciton density at bottleneck and to the condensate density,
as they mainly depend on the phonon coupling rates
for the direct scattering processes. The Coulomb interaction
results both in a modified spectrum (with subsequent
variation of the relaxation rates) and in the occurrence of
coherent scattering processes between the condensate and the
states close to zero momentum.~\cite{leggett01} Both effects are mainly important for the lowest-lying states.
In order to correctly describe
the dominant role played by the low-energy states, the parameter
$\hbar\bar{\omega}$ has to be set to a value lower than a
simple average of the excited energy eigen-values. This parameter represents in our model the energy gap between condensate and excited states. It thus plays the same role as the finite-size energy quantization in a fully two-dimensional system of finite size. We follow this prescription and therefore set the parameter $\hbar\bar{\omega}$ to the energy gap that would result from the system area $A$. The presence of more excited levels would imply an increase of the total coherent outscattering rate with respect to our simplified assumption which therefore sets a lower bound to the effectiveness of the condensate depletion mechanism.}

\section{Numerical solution}
For the numerical evaluation, parameters of a typical AlGaAs
microcavity with one embedded GaAs quantum well have been used,
with a quantization area $A=100~{\mu}\mbox{m}^{2}$. The quantity
$A$ enters the definition of the phonon coupling
terms~\cite{tassone97} as well as the expression for
$\rho_x$~\cite{porras02} and implies a number of excited levels
$\eta=30$,~\cite{porras02} for which we take the representative
energy $\hbar\bar{\omega}=0.1~\mbox{meV}$ (see the discussion in
the previous section). {For these parameters, the
effective phonon coupling strengths linking the bottleneck and
the low-lying states are $g_{c},\tilde{g} \simeq 1~\mu\mbox{eV}$,
resulting in a very long relaxation time ($\tau_{relax}\simeq
100~\mbox{ps}$), consistently with the adiabatic assumption that
we have made.} The Coulomb matrix element is evaluated to be
$\hbar v =5 \times 10^{-4}~\mbox{meV}$.~\cite{ciuti03} The other
parameters are $T=10~\mbox{K}$, $\hbar\gamma_0 = 0.2~\mbox{meV} ,
\hbar\gamma_1 = 1~\mbox{meV}$, $\hbar\Omega_{R}=3.5~\mbox{meV}$,
$\hbar(\omega_{1}-\bar{\omega})=1.9~\mbox{meV}$ and
$\rho_{x}=10^{4}$.~\cite{porras02} The value $\hbar\gamma_0=0.2$
meV is typical for the radiative linewidth of a polariton at zero
exciton-cavity detuning in such a system. The value
$\hbar\gamma_1=1$ meV for the exciton-like part of the polariton
branch, on the other hand, accounts for the density-dependent
exciton dephasing rate~\cite{jahnke96} expected in the vicinity
of the transition density resulting from our calculations. We
have solved numerically the set of equations as a function of
time. For each value of the pump density $f$, we observe a
time-dependent transient followed by a stationary solution for
all the quantities.
\begin{figure}[h!]
\includegraphics[width=.47 \textwidth]{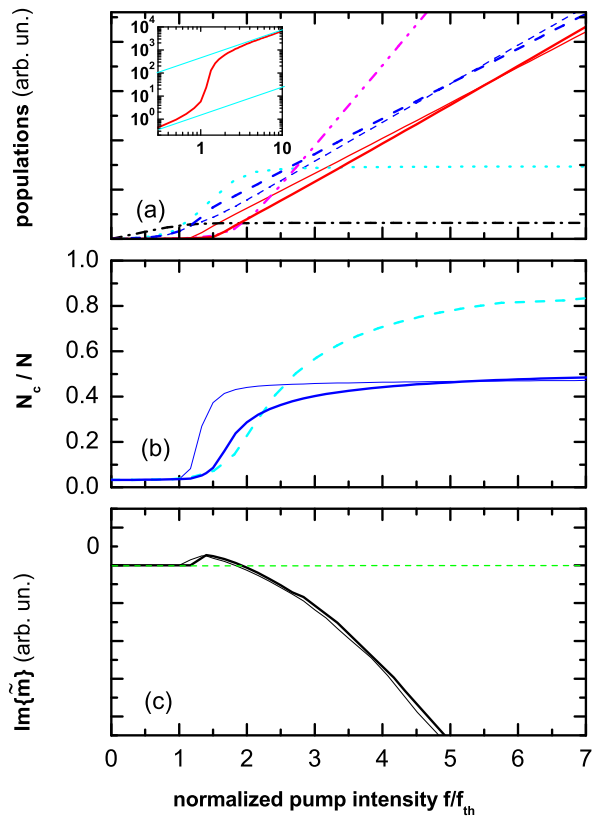}
\caption{\label{fig2}Steady state solutions vs normalized pump
intensity for $\alpha=0.4$ (thin lines) and for $\alpha$
calculated self-consistently (thick lines): detailed discussion in
the text. (a) Solid line: $N_{c}$ (plotted also on a log-log
scale in the inset). Dashed line: $\tilde{N}$. Dot-dashed line:
$N_{1}$. The same quantities $N_{c}$ (dot-dot-dashed) and
$\tilde{N}$ (dotted), computed neglecting the Coulomb
interaction, are also plotted. (b) The corresponding condensate
fraction (solid line) compared to the result computed neglecting
Coulomb interaction (dashed line). (c) The imaginary part of the
Coulomb scattering amplitude $\mbox{Im}\{\tilde{m}\}$.}
\end{figure}
Fig. \ref{fig2}(a) displays the stationary populations as a
function of $f$. {In order to check that the
choice of parameter $\alpha$ is not critical, we compare the
results obtained using a constant value $\alpha=0.4$ with the
results obtained calculating self-consistently $\alpha$ from Eq.
(\ref{eq:alfa_sc}), at each step of the relaxation. The figure clearly shows
that the two methods give very similar results.
Obviously, by using a density-independent $\alpha$,
the condensate fraction at low pump density is overestimated)}. Below
threshold, Boltzmann relaxation results in an increase of the
polariton population, in which the condensate population remains
microscopic, i.e. $N_c\ll\tilde{N}$. A threshold occurs at
$f=f_{th}$, for which $N_{c}\simeq 1$. Above threshold, the
bottleneck population $N_1$ reaches a saturation value while the
condensate population $N_{c}$ becomes a macroscopic fraction of
the non condensed polariton population $\tilde{N}$, which in turn
continues to grow with $f$. The behaviour of the ratio $N_c/N$ is
plotted in Fig. \ref{fig2}(b). At $f\gg f_{th}$ the condensate
fraction approaches a finite value lower than $1$, consistent
with $\alpha=0.4$. Fig. \ref{fig2}(c) displays the imaginary part
of the steady-state Coulomb scattering amplitude $\tilde{m}$.
Close to threshold, in correspondence to low values of the
condensate fraction, this quantity takes positive values, thus
favoring condensation. Above threshold, on the other hand, it
takes large negative values, resulting in condensate depletion.
Hence, Coulomb interaction plays a crucial role during phase
transition, as expected according to both laser \cite{mandel95}
and BEC~\cite{pines66,pita03,griffin95} quantum theories. In
order to clarify the effect of the Coulomb interaction, we
compare in Fig. \ref{fig2}(a) and (b) the steady state solutions
obtained neglecting all Coulomb terms in (\ref{eqall}). By
inspection of Eqs. (\ref{eqall}), it is clear that in this case
the populations obey a standard Boltzmann dynamics. Without
Coulomb interaction, therefore, a standard three-level Boltzmann
equation is recovered. At threshold, the condensate population
starts to grow due to final state stimulation and the system
undergoes a complete transition to a fully condensed regime with
a condensate fraction equal to 1.

\begin{figure}[h!]
\includegraphics[width=.47 \textwidth]{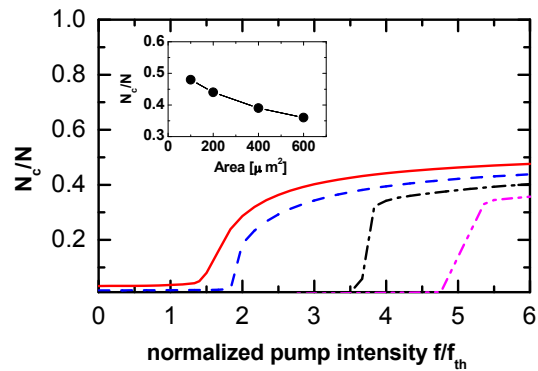}
\caption{\label{fig3}Condensate fraction as a function of the
pump intensity for different values of $A$.
{$A=100~{\mu}\mbox{m}^{2}$ (solid line),
$A=200~{\mu}\mbox{m}^{2}$ (dashed),
$A=400~{\mu}\mbox{m}^{2}$ (dot-dashed) and
$A=600~{\mu}\mbox{m}^{2}$ (dot-dot-dashed). The pump
intensity is renormalized to the threshold value for
$A=100~{\mu}\mbox{m}^{2}$. In the inset we show the asymptotic condensate
fraction above pump threshold, as a function of the system area.}}
\end{figure}

{The threshold $f_{th}$ depends on the energy gap
$\bar{\omega}$. In the
limit of a system of infinite size, this energy gap vanishes and
$f_{th}$ becomes infinite. In order to understand how the size of
the system affects the condensate fraction, on the other hand, in
Fig. \ref{fig3} we compare the results obtained with different
values of $A$. For this comparison, we assume that $\bar{\omega}$
scales as $A^{-1}$, according to the energy quantization in a quantum system of finite size.}
{The results plotted in Fig. \ref{fig3} show that
the condensate fraction reached far above threshold depends on
the system size. This condensate depletion is due to the coherent
scattering terms out of the condensate that originate from the
kinetic HFB equations. In the thermodynamic limit, therefore, our
model predicts a vanishing condensate fraction, as imposed by the
Hohenberg-Mermin-Wagner theorem.\cite{lauwers03} If in the present model we
neglect the Coulomb interaction, the coherent scattering terms
are absent, resulting in a condensate fraction that always tends
to 1 far above the threshold pump intensity (see Fig. \ref{fig2} (b)). The present kinetic model therefore correctly
reproduces the behaviour of a two-dimensional Bose system in the
thermodynamic limit, where the long-wavelength fluctuations are
expected to destroy the condensate.}

\begin{figure}[h!]
\includegraphics[width=.47 \textwidth]{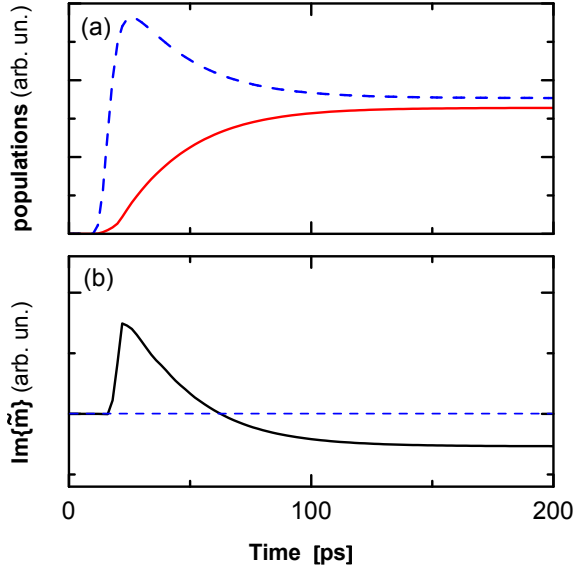}
\caption{\label{fig4}Time evolution of $N_c$ (solid line in (a)),
$\tilde{N}$ (dashed line in (a)) and $\mathop{\rm
Im}\{\tilde{m}\}$ (in (b)) for a pump value $f=2 f_{th}$.}
\end{figure}

To better understand the interplay between scattering amplitudes
and condensate growth, we display the time evolution of $N_c$ and
$\tilde{N}$ in Fig. \ref{fig4}(a) and of the imaginary part of
$\tilde{m}$ in Fig \ref{fig4}(b).
{At short times, the quantity have not a
stationary value, due to the relaxation dynamics and the
condensate population is small.} Correspondingly, the quantity
$\mathop{\rm Im}\{\tilde{m}\}$ takes positive values. At longer
times, when all the quantities reach a stationary value, with a
macroscopic occupation of the condensate, $\mathop{\rm
Im}\{\tilde{m}\}$ has turned to negative values, implying the
resulting condensate depletion.

The Coulomb interaction is therefore the key mechanism
determining the condensate fraction under steady-state
conditions. For the parameters used in Fig. \ref{fig2}, the
saturation value of $N_1$ corresponds here to a bottleneck
polariton density $n_B\sim 2 \times 10^{11}\,\mbox{cm}^{-2}$,
larger than the optical saturation density, suggesting that
condensation cannot occur in a simple system with one quantum
well. When using the parameters of the experiment by Deng {\em et
al},~\cite{deng02,deng03} i.e. considering a sample containing 12
quantum wells at $T=4 K$ with $\hbar\Omega_R=7.5~\mbox{meV}$, we
obtain a qualitatively similar behaviour (not shown) with $N_{1}=
20$ at threshold, resulting in a bottleneck polariton density per
quantum well $n_B\sim 7 \times 10^{9}\,\mbox{cm}^{-2}$, lower than
saturation density and in fairly good agreement with the
experimental estimate.~\cite{deng02,deng03} We argue that, in the
experiment by Deng {\em et al.}, the unusually high values
measured for the two-photon correlation function might be
explained by the coexistence of condensed and non-condensed
phases even far above threshold.

\section{Conclusions}

We have presented a kinetic theory of microcavity polariton
condensation. The theory is based on a number-conserving HFB
description of the polariton quantum field, treated using a
density-matrix formalism. This allows inclusion of various
scattering processes, inducing the energy relaxation and the
final-state bosonic stimulation which causes the condensate to
grow. In particular, we describe the polariton-phonon scattering,
but other scattering mechanisms \cite{porras02,doan05} could in
principle be included. Differently from BEC kinetic models in
atomic physics, here the HFB field dynamics is much faster than
the relaxation kinetics. As a consequence, the steps of the
relaxation kinetics follow adiabatically the spectrum of
collective modes stemming from the HFB field equations. We show
that in this non-equilibrium regime, the condensate fluctuations
strongly influence the phase transition even far above threshold.
In particular, they induce scattering from the condensate to the
excitation modes that can result in a small condensate fraction
in presence of slow energy relaxation rates.

Within a few-level model, we have performed a numerical evaluation of the
condensation kinetics. It turns out that in realistic experimental conditions
the condensate fraction above threshold approaches asymptotically a value
significantly lower than 1, depending on the system size, as expected for a
two-dimensional Bose system. The condensate fraction can even become vanishing
if the size exceeds a few tens of $\mu$m.

We conclude that the coexistence of condensate and
non-condensate, caused by the Coulomb interaction, is a dominant
aspect of the polariton condensation dynamics. This feature,
unpredicted by models based on a standard Boltzmann kinetic
approach, can affect strongly the coherence properties of
condensed polaritons and possibly prevent condensation in the
most common experimental conditions. This result holds great
importance in the light of the numerous experimental claims of
polariton Bose-Einstein condensation
\cite{dang98,senellart99,deng02,deng03,boeuf00,richard05} and of
the recent achievements in lateral confinement of microcavity
polaritons over the micrometric scale.~\cite{dasbach02,deveaud05}

\acknowledgements We are grateful to I. Carusotto and R.
Zimmermann for fruitful discussions. We acknowledge financial
support from the Swiss National Foundation through project N.
620-066060.

\section{Appendix}

The equations of motion appearing in this paper are derived in the
mean-field limit, i.e. factoring higher order correlation terms
in single-particle or Bogolubov quasi-particle population terms.
In this appendix we give the details of this prescription. Let us
consider the system having a defined number of condensed and
non-condensed particles, $N_c$ and $\tilde{N}=\sum \tilde{N}_k$
respectively. We compute the following expectation values of the
two-particle quantities
\begin{equation}
\langle \hat{a}^{\dagger}\hat{a}^{\dagger}\hat{a}\hat{a}\rangle \simeq N_c (N_c - 1)
\end{equation}
and
\begin{equation}
\sum_{kk'qq'}\langle \tilde{p}_k^{\dagger} \tilde{p}_{k'}^{\dagger} \tilde{p}_q \tilde{p}_{q'} \rangle \simeq 2 \tilde{N}^2 - \tilde{N} - \sum_{k} \tilde{N}_k^2.
\end{equation}
Consistently, we can introduce an approximated Hamiltonian in
terms of linearized operators, having the same expectation values
as obtained in mean-field limit. The linearized operators are
\begin{eqnarray}
\hat{a}^{\dagger} \hat{a}^{\dagger}\hat{a}&\simeq&
(N_c-1)\hat{a}^{\dagger}, \\
\hat{a}^{\dagger} \hat{a}\hat{a}& \simeq & N_c \hat{a}, \\
\tilde{p}_k^{\dagger}\tilde{p}_{k'}^{\dagger}\tilde{p}_q
\tilde{p}_{q'} &\simeq &
\tilde{N}_{k,q}\tilde{p}_{k'}^{\dagger}\tilde{p}_{q'}+\tilde{N}_{k,q'}\tilde{p}_{k'}^{\dagger}\tilde{p}_q
\\
& - & \delta_{k,k'} \delta_{k',q} \delta_{q,q'}
(\tilde{N}_k+1)\tilde{p}_k^{\dagger}\tilde{p}_k, \nonumber \\
\tilde{p}_q^{\dagger}\tilde{p}_{-q}^{\dagger}\tilde{p}_k & \simeq
&
(\tilde{N}_{q,k}-\delta_{q,k})\tilde{p}_{-q}^{\dagger}+\tilde{N}_{-q,k}\tilde{p}_{q}^{\dagger},\\
\tilde{p}_{q+q'-k}^{\dagger}\tilde{p}_{q'}\tilde{p}_q &\simeq &
\tilde{N}_{q+q'-k,q'}\tilde{p}_{q}+\tilde{N}_{q+q'-k,q}\tilde{p}_{q'}
\\
& - & \delta_{q+q'-k,q} \delta_{q,q'} \tilde{N}_q\tilde{p}_q, \nonumber \\
\tilde{p}_k^{\dagger}\tilde{p}_{k}^{\dagger}\tilde{p}_q &\simeq &
\tilde{N}_{k}\tilde{p}_{q}+\tilde{N}_{k,q}\tilde{p}_{k}-\delta_{k,q}\tilde{N}_k\tilde{p}_k.
\end{eqnarray}
These relations can be used to factor the correlations entering
the equation for the scattering amplitudes $\tilde{m}_k$
\begin{eqnarray}
\dot{\tilde{m}}_k &=& - 2 i (\omega_k-\omega_0 - 2 v) \tilde{m}_k + v \langle \hat{a}^{\dagger} \hat{a}^{\dagger} (\hat{a}^{\dagger} \hat{a} + \hat{a} \hat{a}^{\dagger}) \tilde{p}_k \tilde{p}_{-k} \rangle \nonumber \\
&-& 4 v \sum_q \langle \hat{a}^{\dagger} \hat{a}^{\dagger} \tilde{p}_q^{\dagger} \tilde{p}_{q} \tilde{p}_k \tilde{p}_{-k} \rangle \nonumber \\
&-& v \sum_q \langle (\hat{a} \hat{a}^{\dagger} + \hat{a}^{\dagger}\hat{a})\tilde{p}_q^{\dagger} \tilde{p}_{-q}^{\dagger} \tilde{p}_k \tilde{p}_{-k} \rangle \nonumber \\
&+& v \langle \hat{a}^{\dagger} \hat{a}^{\dagger} \hat{a} \hat{a} (\tilde{p}_{-k}^{\dagger} \tilde{p}_{-k} + \tilde{p}_k \tilde{p}_{k}^{\dagger}) \rangle \nonumber \\
&+& v \sum_{qq'}\langle \hat{a}^{\dagger} \hat{a}^{\dagger}
(\tilde{p}_{q+q'-k}^{\dagger} \tilde{p}_{-k} + \tilde{p}_k
\tilde{p}_{q+q'+k}^{\dagger}) \tilde{p}_q \tilde{p}_{q'}\rangle.
\end{eqnarray}
In detail, we apply the following factorizations
\begin{equation}
\langle\hat{a}^{\dagger} \hat{a}^{\dagger} \hat{a}^{\dagger}
\hat{a} \tilde{p}_k \tilde{p}_{-k}\rangle \simeq (N_c - 2)
\tilde{m}_k
\end{equation}
\begin{equation}
\langle \hat{a}^{\dagger} \hat{a}^{\dagger} \hat{a}
\hat{a}^{\dagger} \tilde{p}_k \tilde{p}_{-k} \rangle \simeq  (N_c
-1)\tilde{m}_k
\end{equation}
\begin{eqnarray}
\langle\hat{a}^{\dagger}\hat{a}^{\dagger}\tilde{p}_q^{\dagger}\tilde{p}_{q}\tilde{p}_k\tilde{p}_{-k}
\rangle  &\simeq& (\tilde{N}_q-\delta_{q,k}
\tilde{N}_k-\delta_{q,-k}\tilde{N}_{-k})\tilde{m}_k \nonumber \\
&+&\tilde{N}_{q,k}\tilde{m}_{q,-k}+\tilde{N}_{q,-k}\tilde{m}_{q,k}
\end{eqnarray}
\begin{equation}
\langle (\hat{a} \hat{a}^{\dagger} +
\hat{a}^{\dagger}\hat{a})\tilde{p}_q^{\dagger}
\tilde{p}_{-q}^{\dagger} \tilde{p}_k \tilde{p}_{-k} \rangle
\simeq (1+2 N_c) \langle \tilde{p}_q^{\dagger}
\tilde{p}_{-q}^{\dagger} \tilde{p}_k \tilde{p}_{-k} \rangle\\
\end{equation}
\begin{equation}
\langle \hat{a}^{\dagger} \hat{a}^{\dagger} \hat{a} \hat{a}
(\tilde{p}_{-k}^{\dagger} \tilde{p}_{-k} + \tilde{p}_k
\tilde{p}_{k}^{\dagger}) \rangle  \simeq
N_c(N_c-1)(1+\tilde{N}_{-k}+\tilde{N}_k)\\
\end{equation}
\begin{eqnarray}
\langle \hat{a}^{\dagger} \hat{a}^{\dagger}
\tilde{p}_{q+q'-k}^{\dagger} \tilde{p}_{-k} \tilde{p}_q
\tilde{p}_{q'}\rangle  &\simeq & 2\tilde{N}_{q+q'-k,q'}
\tilde{m}_{q,-k} \nonumber \\
&+& \tilde{N}_{q+q'-k,-k} \tilde{m}_{q,q'} \nonumber \\
&-& \delta_{q,k}\delta_{q',k} \tilde{N}_k \tilde{m}_k)
\end{eqnarray}
\begin{eqnarray}
\langle \hat{a}^{\dagger} \hat{a}^{\dagger} \tilde{p}_k
\tilde{p}_{q+q'+k}^{\dagger} \tilde{p}_q \tilde{p}_{q'}\rangle
&\simeq &2\tilde{N}_{q+q'+k,q'}
\tilde{m}_{q,-k} \nonumber \\
&+& \tilde{N}_{q+q'+k,k} \tilde{m}_{q,q'} \nonumber \\
&-& \delta_{-k,q'}\delta_{-k,q} \tilde{N}_{-k} \tilde{m}_k).
\end{eqnarray}
Notice that we can also rewrite the double sums as
\begin{eqnarray}
\sum_{qq'}\tilde{N}_{q+q'-k,q'}\tilde{m}_{q,-k}&=&\sum_{q,q'=k}
\tilde{N}_{q,k}\tilde{m}_{q,-k}+\sum_{q',q=k}\tilde{N}_{q'}\tilde{m}_k
\nonumber
\\
&-& \tilde{N}_k\tilde{m}_k + \sum_{q,q'\neq k}
\tilde{N}_{q+q'-k,q'}\tilde{m}_{q,-k}\nonumber \\
\end{eqnarray}
and the last term can be neglected, within the assumption of a
spatially homogeneous system. In this way, Eq. (\ref{eq:coulomb})
is recovered.

\bibliographystyle{PRSTY}

\end{document}